\documentclass[12pt]{iopart}
\usepackage[dvips]{epsfig}
\usepackage{iopams}
\usepackage{appendix, color}
\begin{document}
\title{Algorithm for generating new explicitly solvable Schr\" odinger type equations}

\author{Nicolae Cotfas$^1$ and Liviu Adrian Cotfas$^2$}
\address{$^1$Faculty of Physics, University of Bucharest,
    PO Box 76 - 54, Post Office 76, 062590 Bucharest, Romania}
\address{$^2$Faculty of Economic Cybernetics, Statistics and Informatics, Academy of Economic Studies,
     6 Piata Romana, 010374 Bucharest, Romania}
\eads{\mailto{ncotfas@yahoo.com}, \mailto{lcotfas@gmail.com}}

\begin{abstract} In this note we present an algorithm to generate new Schr\" odinger type equations explicitly solvable in terms of orthogonal polynomials or associated special functions.
\end{abstract}
\vspace*{1cm}
\maketitle

\section{Introduction}
Many problems in quantum mechanics and
mathematical physics lead to equations of the type
\begin{equation}\label{hypeq}
\sigma (s)y''(s)+\tau (s)y'(s)+\lambda y(s)=0 \label{eq}
\end{equation}
where $\sigma (s)$ and $\tau (s)$ are polynomials of at most second
and first degree, respectively, and $\lambda $ is a constant. 
These equations are usually called {\em equations of hypergeometric
type} \cite{NSU}, and each of them can be reduced to the self-adjoint form 
\begin{equation}
[\sigma (s)\varrho (s)y'(s)]'+\lambda \varrho (s)y(s)=0 
\end{equation}
by choosing a function $\varrho $ such that 
$(\sigma \varrho )'=\tau \varrho $.

The equation (\ref{hypeq}) is usually considered on an interval $(a,b)$,
chosen such that 
\begin{equation}\begin{array}{r}
\sigma (s)>0\qquad {\rm for\ all}\quad s\in (a,b)\\
\varrho (s)>0\qquad {\rm for\ all}\quad s\in (a,b)\\
\lim_{s\rightarrow a}\sigma (s)\varrho (s)
=\lim_{s\rightarrow b}\sigma (s)\varrho (s)=0.
\end{array}
\end{equation}
Since the form of the equation (\ref{eq}) is invariant under a 
change of variable $s\mapsto cs+d$, it is sufficient to analyse the cases
presented in table 1.
Some restrictions are to be imposed to $\alpha $, $\beta $ in
order the interval $(a,b)$ to exist. 

\begin{table}[h]
\caption{The main cases}
\begin{tabular}{ccllc}
\br
$\sigma (s)$ & $\tau (s)$  & $\varrho (s)$ & $\alpha ,\beta $ &  
$(a,b) $\\
\mr
$1$ & $\alpha s\!+\!\beta $ & ${\rm e}^{\alpha s^2/2+\beta s}$ & $\alpha <0$
& $\mathbb{R}$\\
$s$ & $\alpha s\!+\!\beta $ & $s^{\beta -1} {\rm e}^{\alpha s}$ & 
$\alpha <0$, $\beta >0$& $(0,\infty )$\\ 
$1\!-\!s^2$  & $\alpha s\!+\!\beta $ & $(1\!+\!s)^{-(\alpha -\beta )/2-1}
(1\!-\!s)^{-(\alpha +\beta )/2-1}$ & 
$\alpha <\beta <-\alpha $ & $(-1,1)$\\
$s^2\!-\!1$  & $\alpha s\!+\!\beta $ & $(s\!+\!1)^{(\alpha -\beta )/2-1}
(s\!-\!1)^{(\alpha +\beta )/2-1}$ &
$-\beta <\alpha <0$ & $(1,\infty )$\\
$s^2$  & $\alpha s\!+\!\beta $ & $s^{\alpha -2}{\rm e}^{-\beta /s}$ & $\alpha <0$, $\beta >0$ &
$(0,\infty )$\\
$s^2\!+\!1$  & $\alpha s\!+\!\beta $ & $(1\!+\!s^2)^{\alpha /2-1}{\rm e}^{\beta \arctan s}$ & 
$\alpha <0$ & $\mathbb{R}$\\
\br
\end{tabular}
\end{table}

\noindent The equation (\ref{hypeq}) defines an infinite sequence of orthogonal polynomials
in the case $\sigma (s)\in \{ 1,\ s,\ 1-s^2\}$, and a finite one 
in the case $\sigma (s)\in \{ s^2-1,\ s^2,\ s^2+1\}$.
The literature discussing special function theory and its application to mathematical
and theoretical physics is vast, and there are a multitude of different conventions
concerning the definition of functions. 
The table 1 allows 
one to pass in each case from our parameters $\alpha $, $\beta $ to the parameters
used in different approach.

In section 2  we briefly present some results concerning orthogonal polynomials, 
associated special functions, hypergeometric type operators and related Schr\" odinger 
type operators which are needed in section 3. In quantum mechanics there exist potentials, 
called quasi-exactly solvable, for which it is possible to find a finite portion of the energy 
spectrum and the associated eigenfunctions exactly and in closed form \cite{BD,F,Ga,Go,T,U}. 
An algorithm for generating new explicitly  solvable systems and some applications are presented in section 3.

%
\section{Orthogonal polynomials and associated special functions}
Let $\tau (s)=\alpha s+\beta $ be a fixed polynomial, and let
\begin{equation}
\lambda _\ell\!=-\frac{\sigma ''(s)}{2}\ell (\ell -1)-\tau '(s)\ell
\!=-\frac{\sigma ''}{2}\ell (\ell -1)-\alpha \,\ell
\end{equation}
for any $\ell \in \mathbb{N}$. It is well-known \cite{NSU} that for $\lambda =\lambda _\ell $,
the equation (\ref{hypeq}) admits a polynomial solution 
$\Phi _\ell =\Phi _\ell ^{(\alpha ,\beta )}$ of at most $\ell $ degree
\begin{equation} \label{eq3}
\sigma (s) \Phi _\ell  ''+\tau (s) \Phi _\ell '+\lambda _\ell \Phi _\ell =0.
\end{equation}
If the degree of the polynomial $\Phi _\ell $ is $\ell $ then it satisfies the
Rodrigues formula \cite{NSU}
\begin{equation}
\Phi _\ell (s)=\frac{B_\ell }{\varrho (s)}\frac{{\rm d}^\ell }{{\rm d}s^\ell }[\sigma ^\ell (s)\varrho (s)]
\end{equation}
where $B_\ell $ is a constant. Based on the relation 
\begin{equation} 
\begin{array}{l} 
\{ \ \delta \in \mathbb{R}\ |\ 
\lim_{s\rightarrow a}\sigma (s)\varrho (s)s^\delta  
=\lim_{s\rightarrow b}\sigma (s)\varrho (s)s^\delta =0 \ \}\\[2mm]
\mbox{}\qquad \qquad =\left\{
\begin{array}{lll}
[0,\infty ) & {\rm if} & \sigma (s)\in \{ 1,\ s,\ 1-s^2\}\\[2mm] 
[0,-\alpha ) & {\rm if} & \sigma (s)\in \{ s^2-1,\ s^2,\ s^2+1\}
\end{array} \right.  
\end{array} 
\end{equation}
one can prove \cite{NC02,NC04,NSU} that the system of polynomials $\{\Phi _\ell \ |\ \ell <\Lambda \}$, where
\begin{equation}
\Lambda \!=\!\left\{ \begin{array}{lcl}
\infty & {\rm for} & \sigma (s)\in \{ 1,\ s,\ 1-s^2\}\\[2mm]
\frac{1-\alpha }{2} & { \rm for } & 
\sigma (s)\in \{ s^2\!-\!1,\ s^2,\ s^2\!+\!1\} 
\end{array}\right.
\end{equation}
is orthogonal with weight function $\varrho (s)$ in $(a,b)$. This means that
equation (\ref{hypeq}) defines an infinite sequence of orthogonal polynomials
\[ \Phi _0,\ \ \Phi _1,\ \ \Phi _2,\ ... \]
in the case $\sigma (s)\in \{ 1,\ s,\ 1-s^2\}$, and a finite one 
\[ \Phi _0,\ \ \Phi _1,\ \ ...,\ \ \Phi _L \]
with \ $L=\max \{ \ell \in \mathbb{N}\ |\ \ell <(1-\alpha )/2\}$
in the case $\sigma (s)\in \{ s^2-1,\ s^2,\ s^2+1\}$.

The polynomials $\Phi _\ell ^{(\alpha ,\beta )}$ can be expressed (up to a multiplicative constant) in terms of the 
classical orthogonal polynomials as \cite{NC04}
\begin{equation}\label{classical} \fl 
  \Phi _\ell ^{(\alpha ,\beta )}(s)=\left\{ \begin{array}{lcl}
 H_\ell \left(\sqrt{\frac{-\alpha }{2}}\, s-\frac{\beta }{\sqrt{-2\alpha }}\right)  
& {\mbox{}\quad {\rm in\ the\ case}\quad \mbox{}} & \sigma (s)=1\\[2mm]
L_\ell ^{\beta -1}(-\alpha s)  & {\rm in\ the\ case} & \sigma (s)=s\\[2mm]
P_\ell ^{(-(\alpha +\beta )/2-1,\ (-\alpha +\beta )/2-1)}(s)  & {\rm in\ the\ case} & \sigma (s)=1-s^2\\[2mm]
P_\ell ^{((\alpha -\beta )/2-1,\ (\alpha +\beta )/2-1)}(-s)  & {\rm in\ the\ case } & \sigma (s)=s^2-1\\[2mm]
\left(\frac{s}{\beta }\right)^\ell L_\ell ^{1-\alpha -2l}\left(\frac{\beta }{s}\right) 
& {\rm in\ the\ case} & \sigma (s)=s^2\\[2mm]
{\rm i}^\ell P_\ell ^{((\alpha +{\rm i}\beta )/2-1,\ (\alpha -{\rm i}\beta )/2-1)}({\rm i}s) 
& {\rm in\ the\ case} & \sigma (s)=s^2+1
\end{array} \right.
\end{equation}
where $H_\ell $, $L_\ell ^p $ and $P_\ell ^{(p,q)}$ are the Hermite,
Laguerre and Jacobi polynomials, respectively. The relation (\ref{classical}) does not have a very simple form. 
In certain cases we have to consider the classical polynomials 
outside the interval where they are orthogonal or for complex values of parameters.

Let $\ell \!\in \!\mathbb{N}$, $\ell \!<\!\Lambda $, and let $m\!\in \!\{ 0,1,...,\ell \}$.
The functions
\begin{equation}\label{def}
\Phi _{\ell ,m}^{(\alpha ,\beta )}(s)=\kappa ^m(s)\frac{{\rm d}^m}{{\rm d}s^m}\Phi _\ell ^{(\alpha ,\beta )}(s) \qquad {\rm where}\qquad 
\kappa (s)=\sqrt{\sigma (s)}
\end{equation}  
are called the {\em associated special functions}. 
If we  differentiate (\ref{eq3}) $m$ times and then multiply 
the obtained relation by $\kappa ^m(s)$ then we get the equation
\begin{equation}\label{Hm}
\mathcal{H}_m \Phi _{\ell ,m}^{(\alpha ,\beta )}=\lambda _\ell \Phi _{\ell ,m}^{(\alpha ,\beta )}
\end{equation}
where $\mathcal{H}_m$ is the differential operator
\begin{equation} \label{defHm}
\begin{array}{l}
\mathcal{H}_m =-\sigma (s) \frac{d^2}{ds^2}-\tau (s) \frac{d}{ds}
+\frac{m(m-2)}{4}\frac{(\sigma '(s))^2}{\sigma (s)}\\[2mm]   
\mbox{}\qquad \ \ 
 + \frac{m\tau (s)}{2}\frac{\sigma '(s)}{\sigma (s)}
-\frac{1}{2}m(m-2)\sigma ''(s)-m\tau '(s) .
\end{array}
\end{equation}
For each $m\!<\!\Lambda $, the special functions $\Phi _{\ell ,m}^{(\alpha ,\beta )}$ with $m\!\leq \!\ell \!<\!\Lambda $ 
are orthogonal with respect to the scalar product 
\begin{equation} \label{scalarprod}
 \langle f,g\rangle 
=\int_a^b\overline{f(s)}\, g(s)\varrho(s)ds.
\end{equation}

The operators $\mathcal{H}_m$ are directly related to some Schr\" odinger type operators.
If $(a,b)\longrightarrow (a',b'):s\mapsto x=x(s)$ is a differentiable bijective mapping such that
$dx/ds=\pm 1/\kappa (s)$  and $(a',b')\longrightarrow (a,b):x\mapsto s(x)$ is its inverse then the functions
\begin{equation} \label{philm}
\Psi _{\ell ,m}^{(\alpha ,\beta )}(x)=\sqrt{\kappa (s(x))\, \varrho (s(x))}\, \Phi _{\ell ,m}^{(\alpha ,\beta )}(s(x)).
\end{equation}
with $m\leq \ell <\Lambda $ are orthogonal \cite{NC02}
\[ \begin{array}{l}\int_{a'}^{b'}\overline{\Psi }_{\ell ,m}^{(\alpha ,\beta )}(x)\Psi _{k,m}^{(\alpha ,\beta )}(x)dx
=0 \qquad {\rm for}\quad \ell \neq k
\end{array} \]
and satisfy the equation \cite{NC02,NC06}
\begin{equation} \label{Schrodtype}
\left[-\frac{d^2}{dx^2}+V_m(x)\right] \Psi _{\ell ,m}^{(\alpha ,\beta )}=\lambda _\ell  \, \Psi _{\ell ,m}^{(\alpha ,\beta )}
\end{equation}
where $V_m(x)$, defined in terms of the function $\eta (s)=1/\sqrt{\kappa (s)\, \varrho (s)}$,  is
given by 
\begin{equation} \begin{array}{l} 
V_m(x)=\left[ \frac{m(m-2)}{4}\frac{(\sigma '(s))^2}{\sigma (s)}   
 + \frac{m\tau (s)}{2}\frac{\sigma '(s)}{\sigma (s)}-\frac{1}{2}m(m-2)\sigma ''(s)\right.\\[2mm] 
\quad \qquad \qquad \qquad \qquad \left.  -m\tau '(s)\!-\!\sigma (s)\frac{\eta ''(s)}{\eta (s)}
-\tau (s)\frac{\eta '(s)}{\eta (s)}\right]_{s=s(x)}.
\end{array}
\end{equation}
For example, in the case $\sigma (s)\!=\!1$, the change of variable 
$\mathbb{R}\!\rightarrow \!\mathbb{R}:\, x\mapsto s(x)\!=\!x$ leads to
\begin{equation}\label{so}
\begin{array}{l}
V_m(x)=\frac{\alpha ^2}{4}x^2+\frac{\alpha \, \beta }{2}x+\frac{\beta ^2}{4}+\frac{ \alpha }{2}-\alpha m .
\end{array}
\end{equation}

%
%
\section{New explicitly solvable Schr\" odinger type equations}
In the case of a second order differential equation 
\begin{equation} \label{sode}
\begin{array}{l}
\left[A(r)\frac{d^2}{dr^2}+B(r)\frac{d}{dr}+C(r)\right] \psi (r)=0
\end{array}
\end{equation}
with $A(r)\neq 0$ we can eliminate the first order derivative by using the function
\begin{equation}
\begin{array}{l}
h(r)={\rm exp}\left( \int ^r\frac{B(t)}{2A(t)}dt  \right).
\end{array}
\end{equation}
The equation (\ref{sode}) is equivalent to the equation
\begin{equation}
\begin{array}{l}
\frac{1}{A(r)}\, h(r)\left[A(r)\frac{d^2}{dr^2}+B(r)\frac{d}{dr}+C(r)\right]\frac{1}{h(r)}\ h(r) \psi (r) =0
\end{array}
\end{equation}
which can be written as \cite{DW,M}
\begin{equation} 
\begin{array}{l}
\left[\frac{d^2}{dr^2}+\frac{4A(r)\, C(r)-2A(r)\, B'(r)+2B(r)\, A'(r)-B^2(r)}{4A^2(r)}\right]\,  h(r) \psi (r)=0.
\end{array}
\end{equation}

The Schr\" odinger type equations  (\ref{Schrodtype})  have the form \cite{NC06}
\begin{equation} \label{hteq}\fl 
\begin{array}{l}
\left[-\frac{d^2}{dx^2}\!+\!C_1(\alpha ,\beta , m)\, I_1(x)\!+\!C_{-1}(\alpha ,\beta , m)\, I_{-1}(x)\!+\!
C_0(\alpha ,\beta ,m,\ell )\right] \Psi _{\ell ,m}^{(\alpha ,\beta )}(x)\!=\!0.
\end{array}
\end{equation}
If, for $k\!\in \!\{ -1,1\}$, there exists a differentiable bijective mapping 
\[
(\tilde a,\tilde b)\longrightarrow (a',b' ):\, r\mapsto x(r)
\]
such that 
\[
x'(r)=\frac{1}{\sqrt{I_k(x(r))}}
\]
then the  equation (\ref{hteq}) is equivalent to
\begin{equation}
\begin{array}{r}
\left[ -\frac{d^2}{dr^2}\!+\!C_{-k}(\alpha ,\beta ,m)\, \frac{I_{-k}(x(r))}{I_k(x(r))}\!+\!
\frac{C_0(\alpha ,\beta ,m,\ell )}{I_k(x(r))}\!-\!\frac{5}{16}\frac{({I'_k}(x(r)))^2}{(I_k(x(r)))^3}\!\right.\\[3mm]
\left.  +\!\frac{1}{4}\frac{I''_k(x(r))}{(I_k(x(r)))^2}\!-\!E\right]
\sqrt[4]{I_k(x(r))}\, \Psi _{\ell ,m}^{(\alpha ,\beta )}(x(r))\!=\!0
\end{array}
\end{equation}
where $E=-C_k(\alpha ,\beta ,m)$. We know that $\alpha $, $\beta $, $\ell $ and $m$ must satisfy certain restrictions.

In \cite{DW} the authors consider the Schr\"odinger equation (translated harmonic oscillator)
\begin{equation} \label{tho} 
\begin{array}{l}
\left[-\frac{d^2}{dx^2}\!+\!\theta ^2\, x^2+\rho \, x+\lambda \right] \phi (x) \!=\!0.
\end{array}
\end{equation}
In this case the substitution $x=\sqrt{2r}$ leads to the equation 
\begin{equation}\label{dw1}  
\begin{array}{l}
\left[-\frac{d^2}{dr^2}\!+\! \frac{\rho }{\sqrt{2r}}+\frac{\lambda }{2r}-\frac{3}{16}\frac{1}{r^2}+\theta ^2\right] \sqrt[4]{r}\ \phi (\sqrt{2r}) \!=\!0.
\end{array}
\end{equation}
and the substitution $x=\sqrt[3]{(3r/2)^2}$ to the equation 
\begin{equation} \label{dw2} 
\begin{array}{l}
\left[-\frac{d^2}{dr^2}\!+\!\theta ^2\left(\frac{3r}{2}\right)^{\frac{2}{3}}\!+\!\lambda \left(\frac{2}{3r}\right)^{\frac{2}{3}}\!-\!\frac{5}{36}\frac{1}{r^2}\!+\rho  \right] \, \sqrt[6]{r}\ \phi  (\sqrt[3]{(3r/2)^2})\!=\!0.
\end{array}
\end{equation}
In \cite{DW} these equations are considered as two new exactly solvable Schr\" odinger equations.
This is not obvious because, in equation (\ref{tho}), the eigenvalue $\lambda $ depends on $\theta ^2$ and $\rho $. 
The Schr\" odinger type equation corresponding to the potential (\ref{so}) 
\begin{equation} 
\begin{array}{l}
\left[-\frac{d^2}{dx^2}\!+\! \frac{\alpha ^2}{4}x^2+\frac{\alpha \, \beta }{2}x+\frac{\beta ^2}{4}+\frac{ \alpha }{2}-\alpha m+\alpha \ell  \right] \Psi _{\ell,m}^{(\alpha ,\beta )}(x)\!=\!0
\end{array}
\end{equation}
is satisfied for  $\alpha \!<\!0$ and  any $m, \ell \!\in \!\mathbb{Z}$ such that
$0\!\leq \!m\!\leq \!\ell $. It is similar to (\ref{tho}), $I_1(x)\!=\!x^2$, \ $I_{-1}(x)\!=\!x$, \ $C_1(\alpha ,\beta ,m)\!=\!\frac{\alpha ^2}{4}$, \ $C_{-1}(\alpha ,\beta ,m)\!=\!\frac{\alpha \beta }{2}$ and $C_0(\alpha ,\beta , m,\ell )\!=\!\frac{\beta ^2}{4}\!+\!\frac{ \alpha }{2}\!-\!\alpha m\!+\!\alpha \ell $. Using the substitution 
 $x=\sqrt[3]{(3r/2)^2}=\sqrt[3]{\frac{9}{4}}\, r^{\frac{2}{3}}$ we get  the equation
\begin{equation}  \fl
\begin{array}{l}
\left[-\frac{d^2}{dr^2}\!+\!\frac{\alpha ^2}{4}\left(\frac{3r}{2}\right)^{\frac{2}{3}}\!+\!\left(\frac{\beta ^2}{4}\!+\!\frac{ \alpha }{2}\!-\!\alpha m\!+\!\alpha \ell \right) \left(\frac{2}{3r}\right)^{\frac{2}{3}}\!-\!\frac{5}{36}\frac{1}{r^2}\!+ \frac{\alpha \, \beta }{2}  \right] \, \sqrt[6]{r}\ \Psi _{\ell,m}^{(\alpha ,\beta )} (\sqrt[3]{\frac{9}{4}}\, r^{\frac{2}{3}})\!=\!0
\end{array}
\end{equation}
and by using the substitution $x=\sqrt{2r}$ the equation
\begin{equation}\label{so2} \fl  
\begin{array}{l}
\left[-\frac{d^2}{dr^2}\!+\! \frac{\alpha \, \beta }{2}\, \frac{1}{\sqrt{2r}}+\left(\frac{\beta ^2}{4}+\frac{ \alpha }{2}-\alpha m+\alpha \ell \right)\frac{1 }{2r}-\frac{3}{16}\frac{1}{r^2}+\frac{\alpha ^2}{4}\right] \sqrt[4]{r}\ \Psi _{\ell,m}^{(\alpha ,\beta )} (\sqrt{2r}) \!=\!0.
\end{array}
\end{equation}

Let $c_1$ and $c_2$ be two fixed real numbers. The function 
\begin{equation}
\begin{array}{l}
\psi _{\ell ,m}:(0,\infty )\longrightarrow \mathbb{R},\qquad 
\psi _{\ell ,m}(r)=\sqrt[6]{r}\ \Psi _{\ell,m}^{(\alpha ,\beta )} (\sqrt[3]{\frac{9}{4}}\, r^{\frac{2}{3}})
\end{array}
\end{equation}
is a solution of  the equation 
\begin{equation} \label{quantsys}
\begin{array}{l}
\left[-\frac{d^2}{dr^2}\!+\! c_1\, \left(\frac{3r}{2}\right)^{\frac{2}{3}}+c_2\left(\frac{2}{3r}\right)^{\frac{2}{3}}\!-\!\frac{5}{36}\frac{1}{r^2}\right] \psi \!=\!E\psi 
\end{array}
\end{equation}
for $E=-\frac{\alpha \beta }{2}$ if  $\alpha ,\ \beta ,\ \ell ,\ m $ satisfy the system 
\begin{equation}
\left\{ \begin{array}{r}
\frac{\alpha ^2}{4}=c_1\\
\frac{\beta ^2}{4}+\frac{ \alpha }{2}-\alpha m+\alpha \ell =c_2
\end{array} \right.
\end{equation}
and the conditions $m, \ell \!\in \!\{ 0,1,2,...\}$,\ $m\!\leq \!\ell $, \ $\alpha \!<\!0$. In the case $c_1\!\geq \!0$, the functions
\begin{equation}
\begin{array}{l}
\psi _{\ell ,m}^\pm (r)=\sqrt[6]{r}\ 
\Psi _{\ell,m}^{(-2\sqrt{c_1}\, ,\, \pm 2\sqrt{c_2+\sqrt{c_1}(1+2\ell  -2m)})} 
(\sqrt[3]{\frac{9}{4}}\, r^{\frac{2}{3}})
\end{array}
\end{equation}
satisfy the relation
\begin{equation} 
\begin{array}{l}
\left[-\frac{d^2}{dr^2}\!+\! c_1\, \left(\frac{3r}{2}\right)^{\frac{2}{3}}+c_2\left(\frac{2}{3r}\right)^{\frac{2}{3}}\!-\!\frac{5}{36}\frac{1}{r^2}\right] \psi _{\ell ,m}^\pm\!
=\!E_{\ell ,m}^\pm \, \psi _{\ell ,m}^\pm 
\end{array}
\end{equation}
for
\begin{equation}
\begin{array}{l}
E_{\ell ,m}^\pm=\pm 2 \sqrt{c_1\, c_2+c_1\sqrt{c_1}(1+2\ell -2m)}
\end{array}
\end{equation}
if  $m, \ell \!\in \!\{ 0,1,2,...\}$ are such that
\begin{equation}
m\leq \ell \qquad {\rm and}\qquad c_2+\sqrt{c_1}(1+2\ell  -2m)\geq 0.
\end{equation}
By using (\ref{classical}), (\ref{def}),  (\ref{philm}) and table 1, the solutions $\psi _{\ell ,m}^\pm $ can be expressed in terms of Hermite polynomials 
\begin{equation} \fl 
\begin{array}{rl}
\psi _{\ell ,m}^\pm (r) & =\sqrt[6]{r}\ {\rm exp}\left(-\frac{3}{4}\sqrt[3]{\frac{3}{2}}\sqrt{c_1}\, r^{\frac{4}{3}}\pm \sqrt[3]{\frac{9}{4}}\, \sqrt{c_2+\sqrt{c_1}(1+2\ell  -2m)}\, r^{\frac{2}{3}}\right)\\[3mm]
 & \qquad \qquad \times \left[\frac{d^m}{dx^m}H_l\left(\sqrt[4]{c_1}\, x\mp \frac{1}{\sqrt[4]{c_1}}\sqrt{c_2+\sqrt{c_1}(1+2\ell  -2m)}\right)\right]_{x=\sqrt[3]{\frac{9}{4}}\, r^{\frac{2}{3}}}
\end{array}
\end{equation}
and one can remark that they are square integrable on $(0,\infty )$. In view of the well-known relation
$H'_\ell =2\ell \, H_{\ell -1}$, 
the function $\psi _{\ell ,m}^\pm (r)$ coincides up to a multiplicative constant to the function
\begin{equation}  
\begin{array}{rl}
\psi _n^\pm (r) & =\sqrt[6]{r}\ {\rm exp}\left(-\frac{3}{4}\sqrt[3]{\frac{3}{2}}\sqrt{c_1}\, r^{\frac{4}{3}}\pm \sqrt[3]{\frac{9}{4}}\, \sqrt{c_2+\sqrt{c_1}(1+2n)}\, r^{\frac{2}{3}}\right)\\[3mm]
 & \qquad \qquad \times \left[H_n \left(\sqrt[4]{c_1}\, x\mp \frac{1}{\sqrt[4]{c_1}}\sqrt{c_2+\sqrt{c_1}(1+2n)}\right)\right]_{x=\sqrt[3]{\frac{9}{4}}\, r^{\frac{2}{3}}}
\end{array}
\end{equation}
where $n\!=\!\ell \!-\!m$. For $c_1\!\geq \!0$,   the function $\psi _n^\pm (r)$ is an eigenfunction of the Schr\" odinger type operator 
\begin{equation} 
\begin{array}{l}
\mathcal{H}=-\frac{d^2}{dr^2}\!+\! c_1\, \left(\frac{3r}{2}\right)^{\frac{2}{3}}+c_2\left(\frac{2}{3r}\right)^{\frac{2}{3}}\!-\!\frac{5}{36}\frac{1}{r^2}
\end{array}
\end{equation}
corresponding to the eigenvalue
\begin{equation}
\begin{array}{l}
E_n^\pm=\pm 2 \sqrt{c_1\, c_2+c_1\sqrt{c_1}(1+2n)}
\end{array}
\end{equation}
for any $n\!\in \! \{ 0,1,2,...\}$ satisfying the relation
\begin{equation}
n\geq -\frac{c_2}{2\sqrt{c_1}}-\frac{1}{2}.
\end{equation}

Certain solutions of  the equation  (see (\ref{so2}))
\begin{equation}
\begin{array}{l}
\left[-\frac{d^2}{dr^2}\!+\! c_1\, \frac{1}{\sqrt{2r}}+c_2\frac{1 }{2r}-\frac{3}{16}\frac{1}{r^2}\right] \psi \!=\!E\psi 
\end{array}
\end{equation}
may be found by looking for solutions of the system of equations
\begin{equation} \label{syst2}
\left\{ \begin{array}{r}
\frac{\alpha \, \beta }{2}=c_1\\
\frac{\beta ^2}{4}+\frac{ \alpha }{2}-\alpha m+\alpha \ell =c_2
\end{array} \right.
\end{equation}
satisfying the conditions $\alpha \!<\!0$ and  $m, \ell \!\in \!\{ 0,1,2,...\}$ with
$0\leq m\leq \ell $. The corresponding value of $E$ is $-\frac{\alpha ^2}{4}$.
The solution of the system (\ref{syst2}) leads to an equation of third degree, the formulas are more
complicated and will not be presented here. The other cases presented in table 1 may also lead to some
new explicitly solvable systems.

\section{Concluding remarks} 

The method proposed by Derezi\' nski and Wrochna allows us to generate new explicitly solvable systems if we take into consideration the parameter dependence of the eigenvalues of the starting system. Some interesting explicitly solvable systems can be obtained by starting from  solvable hypergeometric type equations containing several parameters.

\mbox{}\\[3mm]
\noindent{{\bf Acknowledgment.}  LAC gratefully acknowledges the financial support provided by ``Doctoral Programme and PhD Students in the education research and innovation
triangle'' under the project POSDRU/6/1.5/S/11. This project is co funded by European Social Fund through The Sectorial Operational Program for Human Resources Development
2007-2013, coordinated by The Bucharest Academy of Economic Studies.
}

\end{document}